\documentclass[twocolumn,           % Format : preprint, twocolumn
               showpacs,            % Pacs : showpacs, noshowpacs
               preprintnumbers,     % Preprint: preprintnumbers,
               aps,                 % Society: ...
               prd,                 % Journal Style : pra, prb, prc, prd, pre,
               a4paper,             % Size : a4paper, ...
               superscriptaddress,  % Affiliation (Title) : groupedaddress,
               nofootinbib,         % Footnote: footinbib, nofootinbib
               tightenlines,        % Remove additional spaces in a line
               floats,floatfix      % Floating pictures and tables
               ]{revtex4}

\usepackage{graphicx}
\usepackage{subfigure}
\usepackage{latexsym}
\usepackage{amsmath,amssymb}        % amssymb includes amsfonts
\usepackage[draft=false]{hyperref}

\input colordvi

\begin{document}

\title{Intermediate inflation in light of the three-year WMAP
observations}   
\author{John D.~Barrow}
\affiliation{DAMTP, Centre for Mathematical Sciences,
Cambridge University, Wilberforce Road, Cambridge CB3 0WA, UK}
\author{Andrew R.~Liddle}
\affiliation{Astronomy Centre, University of Sussex, Brighton BN1 9QH,
UK}
\author{C\'edric Pahud}
\affiliation{Astronomy Centre, University of Sussex, Brighton BN1 9QH,
UK}
\date{\today}
\pacs{98.80.-k \hfill astro-ph/0610807}
\preprint{astro-ph/0610807}

\begin{abstract}
The three-year observations from the Wilkinson Microwave Anisotropy
Probe have been hailed as giving the first clear indication of a
spectral index $n_{\mathrm{s}}<1$. We point out that the data are
equally well explained by retaining the assumption $n_{\mathrm{s}}=1$
and allowing the tensor-to-scalar ratio $r$ to be non-zero. The
combination $n_{\mathrm{s}}=1$ and $r>0$ is given (within the
slow-roll approximation) by a version of the intermediate inflation
model with expansion rate $H(t)\propto \ t^{-1/3}$.  We assess the
status of this model in light of the WMAP3 data.
\end{abstract}

\maketitle

\section{Introduction}

The most striking result from the three-year Wilkinson Microwave
Anisotropy Probe (WMAP) observations \cite{wmap3} is the pressure that
they impose on the Harrison--Zel'dovich spectrum of density
perturbations, for which adiabatic perturbations have the
scale-invariant spectral index $n_{\mathrm{s}}=1$. This spectrum was
first proposed by Harrison \cite{harr} and Zel'dovich \cite{zeld}
because it has metric potential perturbations of the same amplitude on
all scales. This allows small perturbation theory to hold on large and
small scales, and would also allow primordial black-hole formation to
occur over a wide range of mass scales if the amplitude of
fluctuations was sufficiently large \cite{CH}. Harrison--Zel'dovich
spectra arise in pure de Sitter inflationary universe models, but they
\ have also been shown to arise from different non-inflationary
cosmological situations \cite{branpeter}.

The simplest versions of inflation, in which a finite period of de
Sitter-like inflationary expansion occurs, naturally create such a
spectrum of fluctuations because the dynamics have no preferred moment
of time in de Sitter spacetime: an irregularity spectrum with
identical metric perturbations on each scale respects this
invariance. However, there are many variants of inflation for which
the expansion dynamics are not of de Sitter form, and they predict
different spectra of fluctuations; hence it is important to determine
which (if any) of them are consistent with the current observational
data. If adiabatic density perturbations are the only perturbations
present, then the original WMAP3 parameter-estimation analysis
suggested that the Harrison--Zel'dovich spectrum is excluded at quite
high significance \cite{wmap3}. This significance has been reduced by
re-analysis of the inflationary constraints by the WMAP team
(available at Ref.~\cite{wmapanal}), from the viewpoint of the more
sophisticated statistical approach of model selection \cite{PML}, and
by recent papers highlighting possible systematic effects
\cite{syspapers}, but it is nevertheless timely to explore possible
interpretations of these data.

The conclusion that $n_{\mathrm{s}}=1$ is disfavoured by the data is
restricted to the case where adiabatic scalar perturbations are the
only ones present. The best-motivated generalization is the inclusion
of tensor perturbations, which are predicted to be present at some
level by inflation, and parametrized by the tensor-to-scalar ratio
$r$. This is explored in some detail by the WMAP team \cite{wmap3},
and in subsequent papers \cite{nrcons}, with the
conclusion that $n_{\mathrm{s}}\geq 1$ is readily allowed provided
that the value of $r$ is significantly non-zero.

In this \emph{Brief Report}, we analyze a particular class of
inflationary models which give this behaviour, the \emph{intermediate
inflation} model discussed in Refs.~\cite{inter, mus, rendall}. This
was originally introduced as an exact inflationary solution for a
particular scalar field potential, but is perhaps best-motivated as
the slow-roll solution to potentials which are asymptotically of
inverse power-law type, $V\propto \phi^{-\beta}$. This type of
potential is in common use in quintessence models \cite{RPmodel}, but
it also gives viable inflationary solutions, although with this
precise potential form inflation is everlasting and a mechanism has to
be introduced to bring inflation to an end. It also arises in a range
of scalar--tensor gravity theories \cite{maeda}.

As shown by Barrow and Liddle \cite{BL}, the intermediate inflationary
model, in the slow-roll approximation, gives $n_{\mathrm{s}}=1$ to
first order provided $\beta =2$ (see also Ref.~\cite{valli} for a more
extensive discussion of the inflationary generation of the
Harrison--Zel'dovich spectrum, and Ref.~\cite{starob} for the
construction of exact potentials giving $n_{{\rm S}} = 1$ without
slow-roll approximation). In this case, $r$ depends significantly on
scale, falling in value with time and hence becoming smaller on
shorter length-scales. There will be an observable effect provided
inflation ends swiftly enough, so that $r$ was still important at the
horizon crossing of observable perturbations.  More generally, if
$\beta \neq 2,$ the spectral index may exhibit running, approaching
unity at late times; see also the review of this situation in
Ref.~\cite{nrcons}.

\section{Predictions of the model}

A generalization of the intermediate inflation model \cite{inter} used
in the earlier study of Ref.~\cite{BL} has an expansion scale factor
given by (with appropriate choice of time coordinate)
\begin{eqnarray}
&&a(t)=\exp \left( At^{f}+Bt\right) \,, \\
&& \qquad\quad 0<f<1,\quad A>0,~B\geq
0,\;\mathrm{constants}\,.  \label{a} \nonumber
\end{eqnarray}
This is an exact solution of the Friedmann equations ($8\pi G=c=\hbar
=1$) for a flat universe containing a scalar field $\phi (t)$ with
potential $ V(\phi )$, where
\begin{widetext}
\begin{eqnarray}
\phi  &=&\phi _{0}+\left[ \frac{8A(1-f)}{f}\right] ^{1/2}\;t^{f/2}\,,
\label{ex1} \\
V(\phi ) &=&3\left\{ B+Af\left[ \frac{f}{8A(1-f)}\right] ^{(f-1)/f}(\phi
-\phi _{0})^{2(f-1)/f}\right\} ^{2} \label{ex2} %\\
%&& \quad 
-Af(1-f)\left[ \frac{f}{8A(1-f)}\right] ^{(f-2)/f}(\phi -\phi
_{0})^{2(f-2)/f}\,. \; \;
\end{eqnarray}
\end{widetext}
It can be obtained using the solution-generating method of
Ref.~\cite{pb}.  Without loss of generality $\phi _{0}$ can be taken
to be zero. We will now specialise to the pure intermediate
inflationary model of Refs.~\cite{inter,BL} with $B=0$ and $A>0$.

In the slow-roll approximation with $B=0$, the first term on the RHS
of Eq.~(\ref{ex2}) dominates $V$ at large $\phi $, and we have
\begin{eqnarray}
\phi  &=&(2A\beta t^{f})^{1/2}\quad \mathrm{where}\;\;\beta \equiv
4(f^{-1}-1)\,; \\
V(\phi ) &=&\frac{48A^{2}}{(\beta +4)^{2}}(2A\beta )^{\beta /2}\phi
^{-\beta}\,,  \label{slow2}
\end{eqnarray}
as the scalar field rolls down a power-law potential. The first two
slow-roll parameters are then given, in the Hamilton--Jacobi formalism
\cite{LPB}, by
\begin{equation}
\epsilon =\frac{\beta ^{2}}{2\phi ^{2}},\quad \eta =\beta \left( 1+\frac{
\beta }{2}\right) \frac{1}{\phi ^{2}}\,.
\end{equation}
So, the condition for inflation to occur ($\epsilon <1$) is only satisfied
when $\phi ^{2}>\beta ^{2}/2$.

\subsection{First-order considerations}

In order to confront this model with observations, we need to consider
the contribution of the scalar and tensor perturbations which can be
represented by $n_{\mathrm{s}}$ and $r$, respectively. They are
expressed in terms of the slow-roll parameters to first order by
\cite{LL}
\begin{eqnarray}
n_{\mathrm{s}} &=&1-4\epsilon +2\eta =1-\frac{\beta (\beta -2)}{\phi^{2}}
\label{n} \\
r &=&16\epsilon =\frac{8\beta ^{2}}{\phi ^{2}}  \label{r}
\end{eqnarray}
We clearly see that $n_{\mathrm{s}}=1$ and $r>0$ is possible, provided
$\beta =2$. This is the case where $\eta =2\varepsilon $. We see that
an exact scale-invariant spectrum can be obtained to leading order in
slow-roll by both the de Sitter expansion, i.e.~with $a(t)=\exp
(H_{0}t)$ and $H_{0}$ constant, and by the special intermediate
inflationary dynamics with $a(t)=\exp (At^{2/3})$.

\begin{figure}[t]
\centering   %\includegraphics[width=8cm]{nr.eps}
\includegraphics[width=8cm]{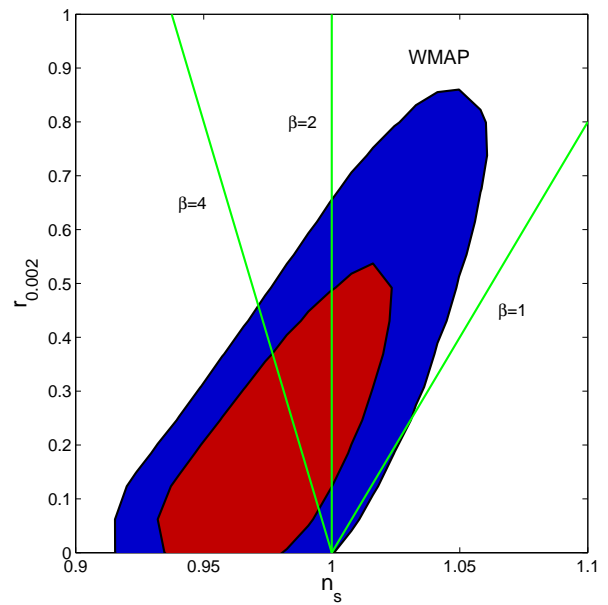}
\caption{Trajectories for different values of the parameter
$\protect\beta$ in the $n_{\mathrm{s}}$--$r$ plane, to first-order in
slow roll. From left to right $\protect\beta =4$, 2, 1. The two
contours correspond to the 68\% and 95\% levels of confidence. The
observational data is from the WMAP analysis at Ref.~\cite{wmapanal},
which updates that of version 1 of Ref.~\cite{wmap3}. The
observational dataset used is WMAP alone, applied to the lcdm+tens
model (without spectral index running).}
\label{f:plot}
\end{figure}

Returning to the general case ($0<f< 1$), this model can be compared
with observations, shown in Fig.~\ref{f:plot}. The relation between
the scalar and tensor perturbations is
\begin{equation}
n_{\mathrm{s}}=1-\frac{\beta -2}{8\beta }r.
\end{equation}
On one hand, we see that $\beta >2$ is well supported by the data,
while on the other, we see that $\beta <2$ allows $n_{\mathrm{s}}>1$,
but becomes rapidly disfavoured when $\beta $ approaches $1$.

In order for our comparison of the intermediate inflationary model's
predictions with the observations to be complete, we must also
consider the time spent by the field in the region of the
$n_{\mathrm{s}}$--$r$ plane allowed by the data. The number of
$e$-foldings between two different values $\phi_{1}$ and $\phi_{2}$ of
the scalar field is given by \cite{BL}
\begin{equation}
N(\phi _{1},\phi _{2})=\frac{1}{2\beta }(\phi _{2}^{2}-\phi _{1}^{2})\,.
\end{equation}
If we assume that inflation begins at the earliest possible stage,
that is at $\phi _{1}^{2}=\beta ^{2}/2$, then Eqs.~(\ref{n}) and
(\ref{r}) can be re-expressed in terms of the number of $e$-foldings,
$N_{b}$, which have passed since the beginning of the inflationary
period:
\begin{eqnarray}
n_{\mathrm{s}} &=&1-\frac{2\beta -4}{4N_{b}+\beta },  \label{nN} \\
r &=&\frac{16\beta }{4N_{b}+\beta }.  \label{rN}
\end{eqnarray}
If we consider a Harrison--Zel'dovich model ($\beta =2$) with the
inclusion of gravitational waves, then we see in Fig.~\ref{f:plot}
that the curve $r=r(n_{\mathrm{s}})$ enters the 95\% confidence region
for $r=0.66$ which corresponds to $N_{b}\simeq 12$. Since the point
$(n_{\mathrm{s}}=1$, $r=0)$ lies just inside the two-dimensional 95\%
confidence contour, the model is viable for all larger values of
$N_{b}$.

\subsection{Second-order corrections}

Next, we show that the second-order corrections to our analysis at
first-order in slow roll are small, and can be neglected to a very
good approximation. Generalizing Eq.~(\ref{n}) in terms of the
slow-roll parameters to second-order, we have \cite{LL,SL}
\begin{equation}
n_{\mathrm{s}}-1=2\eta -4\epsilon -[8(1+C)\epsilon
  ^{2}-(6+10C)\epsilon \eta +2C\xi ^{2}],
\end{equation}
\newline
with $C=-0.73$ a known numerical constant, and $\xi ^{2}(\phi) \equiv
\epsilon \eta -(2\epsilon )^{1/2}\, d\eta/d\phi$. Putting $\beta =2$
(so that we have $n_{\mathrm{s}}=1$ exactly to first order) in the
above expression, we get the second-order correction to the spectral
index:
\begin{equation}
n_{\mathrm{s}}-1=4\epsilon ^{2}.
\end{equation}
Finally, knowing that $r=16\epsilon +O(\epsilon ^{2})$, we obtain to
second order that
\begin{equation}
n_{\mathrm{s}}-1=\frac{r^{2}}{64}.
\end{equation}

The above calculation, which uses the exact solution, corresponds to
the full potential Eq.~(\ref{ex2}). While in the full slow-roll
approximation this gives the same result as the single power-law model,
Eq.~(\ref{slow2}), at second-order the potentials yield different
results. A similar calculation to the above, but using the
$V$-slow-roll approximation \cite{LPB}, shows that the denominator 64
is modified to $384/7$ in that case.

Observations \cite{wmap3} constrain $r$ to be less than $0.65$ (at
95\% confidence). So, for either potential, this extra contribution in
the case with $\beta =2$ is quite negligible once the field enters the
region allowed by the data.

\subsection{The running of the spectral index}

The running of the spectral index in inflationary models is given, to
lowest order in slow-roll, by \cite{LL}
\begin{equation}
\frac{dn_{\mathrm{s}}}{d\ln k}=-8\epsilon ^{2}+10\epsilon \eta -2\xi ^{2}= 
\frac{2\beta ^{2}(\beta -2)}{\phi ^{4}}.  \label{run1}
\end{equation}
Moreover, to lowest order $n_{\mathrm{s}}-1=-\beta (\beta -2)\phi
^{-2}$, which allows us to rewrite this relation as
\begin{equation}
\frac{dn_{\mathrm{s}}}{d\ln k}=\frac{2}{\beta -2}(n_{\mathrm{s}}-1)^{2}.
\label{run2}
\end{equation}
We can deduce from this relation that $\beta =2$ implies no running of
the spectral index to first-order, which was already obvious from the
comment following Eq.~(\ref{n}).

Models with $\beta >2$ feature positive running, which the WMAP3 data
disfavor \cite{wmap3,wmapanal}. However, within the allowed region the
predicted running is very small (for example, it is always less than
0.001 for the $\beta =4$ case), and it would be premature to claim
that the running constraint adds any value to the
$n_{\mathrm{s}}$--$r$ constraints for these models.

\section{Conclusions}

The intermediate inflation model is a viable example of a model with
$n_{\mathrm{s}}=1$ which is permitted by the observational data, due
to the non-zero tensor contribution. In this model, $r$ is
scale-dependent, and we have shown that a good fit to the WMAP3
observations is possible provided observable scales crossed outside
the horizon at least 12 $e$-foldings after the earliest possible
starting-point for inflation.  Arranging this requires that whatever
mechanism is introduced to bring inflation to an end does so with
$\phi > 14$ in reduced Planck units, considering that a minimum of
perhaps 50 $e$-foldings is required to push the perturbations to
observable scales \cite{LL}.

This model serves as a useful phenomenological illustration, in the
light of WMAP3 data, of a type of simple slowly-rolling scalar field
evolution that does not display pure de Sitter inflationary expansion,
but can still produce a Harrison--Zel'dovich spectrum.

For the more general intermediate inflation case with $\beta \neq 2$,
observations constrain $\beta$ to be greater than about one, unless we
are in the regime very close to the Harrison--Zel'dovich
limit. Constraints from running do not presently add extra
information.

\section*{Acknowledgments}

A.R.L.\ was supported by PPARC and C.P.\ in part by the Swiss Sunburst
Fund.  A.R.L.\ acknowledges the hospitality of the Institute for
Astronomy, University of Hawai`i, while this work was being
completed. C.P.\ acknowledges the hospitality of Caltech and Marc
Kamionkowski while this work was completed, during a visit supported by
the Royal Astronomical Society and by Caltech, and thanks Pia Mukherjee
for discussions and advice.

%%%%%%%%%%%%%%%%%%%%%%%%%%%%%%%%%%%%%%%%%%%%%%%%%%%%%%%%%%%%%%%%%%%%%%%%
\end{document}